\documentclass[twocolumn,groupedaddress,showpacs,aip,reprint]{revtex4-1}
\usepackage{graphicx,amsmath}
\usepackage{amssymb}
\usepackage{sidecap}
\usepackage{hyperref}
\usepackage{color}
\usepackage{mhchem}
\begin{document}
\title{Dynamics of tungsten hexacarbonyl, dicobalt octacarbonyl and
their fragments adsorbed on silica surfaces}

\author{Kaliappan Muthukumar}
\affiliation{Institut f\"ur Theoretische Physik, Goethe-Universit\"at Frankfurt, Max-von-Laue-Stra{\ss}e 1, 60438 Frankfurt am Main, Germany}
\author{Roser Valent\'\i}
\affiliation{Institut f\"ur Theoretische Physik, Goethe-Universit\"at Frankfurt, Max-von-Laue-Stra{\ss}e 1, 60438 Frankfurt am Main, Germany}
\author{Harald O. Jeschke}
\affiliation{Institut f\"ur Theoretische Physik, Goethe-Universit\"at Frankfurt, Max-von-Laue-Stra{\ss}e 1, 60438 Frankfurt am Main, Germany}
\email{jeschke@itp.uni-frankfurt.de}

%%%%%%%%%%%%%%%%%%%%%%%%%%%%%%%%%%%%%%%%%%%%%%%%%%%%%%%%%%%%%%%%%%%%%%%%%%%%%%%%%%%%%%%%%%%%%%%%%%%%%%%%%%%%%%%%%%%%
\date{\today}
%%%%%%%%%%%%%%%%%%%%%%%%%%%%%%%%%%%%%%%%%%%%%%%%%%%%%%%%%%%%%%%%%%%%%%%%%%%%%%%%%%%%%%%%%%%%%%%%%%%%%%%%%%%%%%%%%%%%
\newcommand{\wc} {W(CO)$_6$}
\newcommand{\wcfive}{W(CO)$_5$}
\newcommand{\crfive}{Cr(CO)$_5$}
\newcommand{\mofive}{Mo(CO)$_5$}
\newcommand{\mcfive}{M(CO)$_5$}
\newcommand{\mc}{M(CO)$_6$}
\newcommand{\bc}{${\beta}${-}cristobalite}
\newcommand{\sio}{SiO$_2$}
\newcommand{\tio}{TiO$_2$}
\newcommand{\FOH}{fully hydroxylated}
\newcommand{\POH}{partially hydroxylated}
\newcommand{\plat}{CH$_3$(C$_5$H$_5$)Pt[CH$_3$]$_3$}
\newcommand{\efermi}{E$_f$}
\newcommand{\cfourv}{C$_{4v}$}
\newcommand{\ctwov}{C$_{2v}$}
\newcommand{\cthreev}{C$_{vv}$}
\newcommand{\fc}{Fe(CO)$_6$}
\newcommand{\coco}{Co$_2$(CO)$_8$}
\newcommand{\cothree}{Co(CO)$_3$}
\newcommand{\cofour}{Co(CO)$_4$}
\newcommand{\cosix}{Co$_4$(CO)$_6$}
\newcommand{\cotwelve}{Co$_4$(CO)$_{12}$}
\newcommand{\cosixteen}{Co$_4$(CO)$_{16}$}
\newcommand{\alumina}{$\gamma$$-${Al$_2$O$_{3}$}}
%%%%%%%%%%%%%%%%%%%%%%%%%%%%%%%%%%%%%%%%%%%%%%%%%%%%%%%%%%%%%%%%%%%%%%%%%%%%%%%%%%%%%%%%%%%%%%%%%%%%%%%%%%%%%%%%%%%%
\begin{abstract}
  Tungsten and cobalt carbonyls adsorbed on a substrate are typical
  starting points for the electron beam induced deposition of tungsten
  or cobalt based metallic nanostructures. We employ first principles
  molecular dynamics simulations to investigate the dynamics and
  vibrational spectra of {\wc} and {\wcfive} as well as {\coco} and
  {\cofour} precursor molecules on fully and partially hydroxylated
  silica surfaces. Such surfaces resemble the initial conditions of
  electron beam induced growth processes.  We find that both {\wc} and
  {\coco} are stable at room temperature and mobile on a silica
  surface saturated with hydroxyl groups (OH), moving up to half an
  Angstr\"om per picosecond. In contrast, chemisorbed {\wcfive} or
  {\cofour} ions at room temperature do not change their binding site.
  These results contribute to gaining fundamental insight into how the
  molecules behave in the simulated time window of 20~ps and our
  determined vibrational spectra of all species provide signatures for
  experimentally distinguishing the form in which precursors cover a
  substrate.
\end{abstract}

%68.35.-p 	Solid surfaces and solid-solid interfaces: structure and energetics
%68.35.Np 	Adhesion
%68.43.-h 	Chemisorption/physisorption: adsorbates on surfaces
%68.43.Fg 	Adsorbate structure (binding sites, geometry)
%71.15.Nc 	Total energy and cohesive energy calculations
%71.15.Mb 	Density functional theory, local density approximation, gradient and other corrections
%73.20.At 	Surface states, band structure, electron density of states

%%%%%%%%%%%%%%%%%%%%%%%%%%%%%%%%%%%%%%%%%%%%%%%%%%%%%%%%%%%%%%%%%%%%%%%%%%%%%%%%%%%%%%%%%%%%%%%%%%%%%%%%%%%%%%%%%%%%
\maketitle
%%%%%%%%%%%%%%%%%%%%%%%%%%%%%%%%%%%%%%%%%%%%%%%%%%%%%%%%%%%%%%%%%%%%%%%%%%%%%%%%%%%%%%%%%%%%%%%%%%%%%%%%%%%%%%%%%%%%
\section{Introduction\label{Introduction}}
%%%%%%%%%%%%%%%%%%%%%%%%%%%%%%%%%%%%%%%%%%%%%%%%%%%%%%%%%%%%%%%%%%%%%%%%%%%%%%%%%%%%%%%%%%%%%%%%%%%%%%%%%%%%%%%%%%%%
The study of the energetics and dynamics of individual molecules
adsorbed on a substrate is relevant for several prominent fields of
research such as molecular electronics~\cite{Joachim2000,Heath2009},
molecular magnetism~\cite{Miller2000} and
catalysis~\cite{Besenbacher2007}. The possibility to manipulate
individual molecules adsorbed on surfaces using the tip of a scanning
tunneling microscope (STM) leads to an increasing need for theoretical
information of the geometries as well as adsorption and desorption
mechanisms and reaction pathways of adsorbates~\cite{Hla2003}. STM
probes can not only be used to arrange complex molecules on surfaces
but also to measure their vibrational
spectra~\cite{Otero2006}. Molecules adsorbed on insulating surfaces
which due to insufficient conductivity are hard to study by STM
techniques have been investigated successfully using non-contact
atomic force microscopy~\cite{Schwarz2013}.

Here, we will study molecules and substrates that are important in the
context of electron beam induced deposition (EBID) of organometallic
precursor molecules. This is a widely used method to grow size- and
shape-controlled nanometer-sized
structures.~\cite{frabboni:213116,6323037,Wnuk2011257,Randolph2006,ebidsilvis,0957-4484-20-19-195302,utke:1197}
The obtained EBID deposits nevertheless possess a significant
percentage of organic contaminants mainly from carbon and oxygen,
lowering the conductivity of these deposits and thus limiting the
possible applications of EBID.\cite{1367-2630-14-11-113028,
  arXiv:1303.1739,1367-2630-11-3-033032,Botman20081139,li:023130,barry:3165,0957-4484-20-19-195301}
Several pre- and post-fabrication approaches have been employed to
remove these contaminants, but reproducibility is still an issue.
First principles calculations can provide a detailed description of
the microscopic behavior of the deposits and are often used to improve
the quality of the deposition processes. Important progress has been
done, for instance, on atomic layer
deposition~\cite{widjaja:3304,doi:10.1021/cr900056b,jeloaica:542,ald_electrohem,doi:10.1021/cm035009p}
and chemical vapor deposition~\cite{cvd_henry1,cvd_henry2} processes.

Recently~\cite{juan2012,PhysRevB.84.205442,Muthukumar2012}, in an
attempt to understand the EBID growth process, we analyzed by means of
density functional theory (DFT) calculations the interaction of
precursors like {\wc}, {\coco} and {\plat} on two different {\sio}
surfaces (fully and partially hydroxylated) as a representative for
untreated and pretreated EBID surfaces (Hereafter, surfaces
corresponds to \sio\ surfaces unless otherwise mentioned). These
studies illustrate the preferred orientation of the adsorbate and the
nature of the interaction between the precursor molecules and the
{\sio} substrates.  Further, interesting phenomena such as the
spontaneous fragmentation of the carbonyl precursors ({\wc} to
{\wcfive} and {\coco} to two {\cofour} molecules) on the {\POH}
surfaces that correspond to pre-treated surfaces were
observed.\cite{PhysRevB.84.205442,Muthukumar2012}

It has recently been reported that the surface residence time of an
organometallic precursor should be sufficiently long (lasting from
microseconds to milliseconds) to have an efficient deposition
yield.~\cite{1367-2630-11-3-033032} In order to understand these
observations and to improve the conditions for the adhesion of
precursor molecules to the substrate, it is essential to have
knowledge on the behavior of the precursor molecules and their
fragments on the surface \sio\ substrates.  Theoretical and
experimental studies focussing on the adsorption process of free CO on
various surfaces and several molecules on different {\sio} substrates
have been
reported.\cite{philipsen1997relativistic,steininger1982adsorption,chin2006,doi:10.1021/jp301197s}
However, little is known on the dynamics of {\wc} and {\coco}
precursors and their fragments on {\sio} surfaces.  Therefore, in this
work we use first principles molecular dynamics simulations to
investigate the nature of {\wc} and {\coco} molecules adsorption on
fully and partially hydroxylated {\sio} surfaces and provide
quantitative microscopic insight into the stability of these
fragmented precursors on these surfaces and on their vibrational
spectra.

%%%%%%%%%%%%%%%%%%%%%%%%%%%%%%%%%%%%%%%%%%%%%%%%%%%%%%%%%%%%%%%%%%%%%%%%%%%%%%%%%%%%%%%%%%%%%%%%%%%%%%%%%%%%%%%%%%%%
\begin{figure*}
\includegraphics[width=1\textwidth]{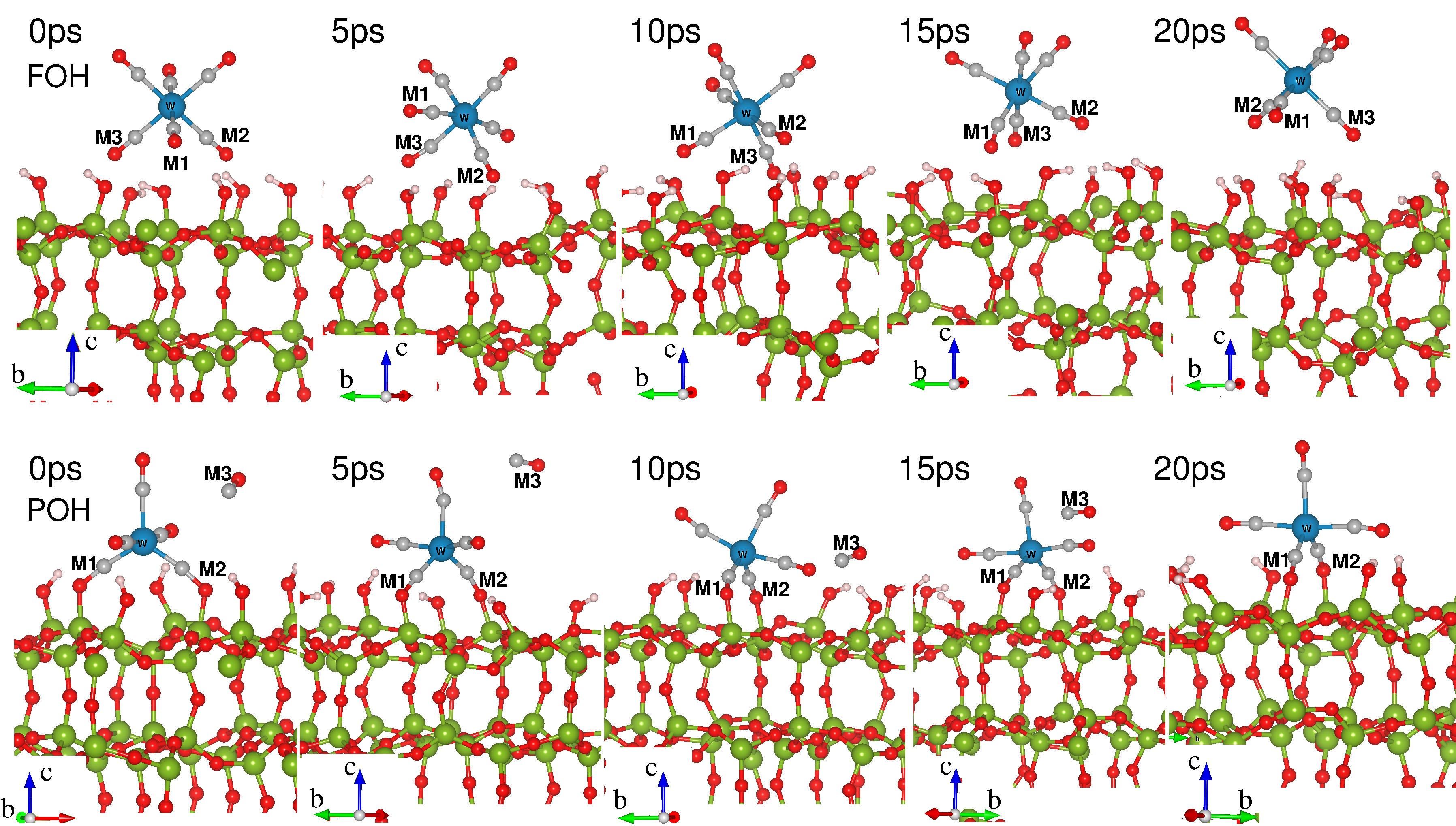}
\caption{(Color online) Adsorption of {\wc} on {\FOH} (FOH, upper
  panel) and {\POH} (POH, lower panel) substrates.  The illustrated
  snapshots are the configurations taken at every 5~ps interval.
  Color code : Green - Si, Red - O, light Blue - W, Gray - C and
  Magenta - H throughout this manuscript.
  M1,M2 and M3 are the labels for specific ligands to identify the
  changes in 2D view and the snapshots have slightly
  different orientations (note the coordinate axis) to better display the changes happening
  to the system}
\label{fig:snapshot}
\end{figure*}
%%%%%%%%%%%%%%%%%%%%%%%%%%%%%%%%%%%%%%%%%%%%%%%%%%%%%%%%%%%%%%%%%%%%%%%%%%%%%%%%%%%%%%%%%%%%%%%%%%%%%%%%%%%%%%%%%%%%

%%%%%%%%%%%%%%%%%%%%%%%%%%%%%%%%%%%%%%%%%%%%%%%%%%%%%%%%%%%%%%%%%%%%%%%%%%%%%%%%%%%%%%%%%%%%%%%%%%%%%%%%%%%%%%%%%%%%
\begin{figure}
\centering
\includegraphics[width=0.47\textwidth]{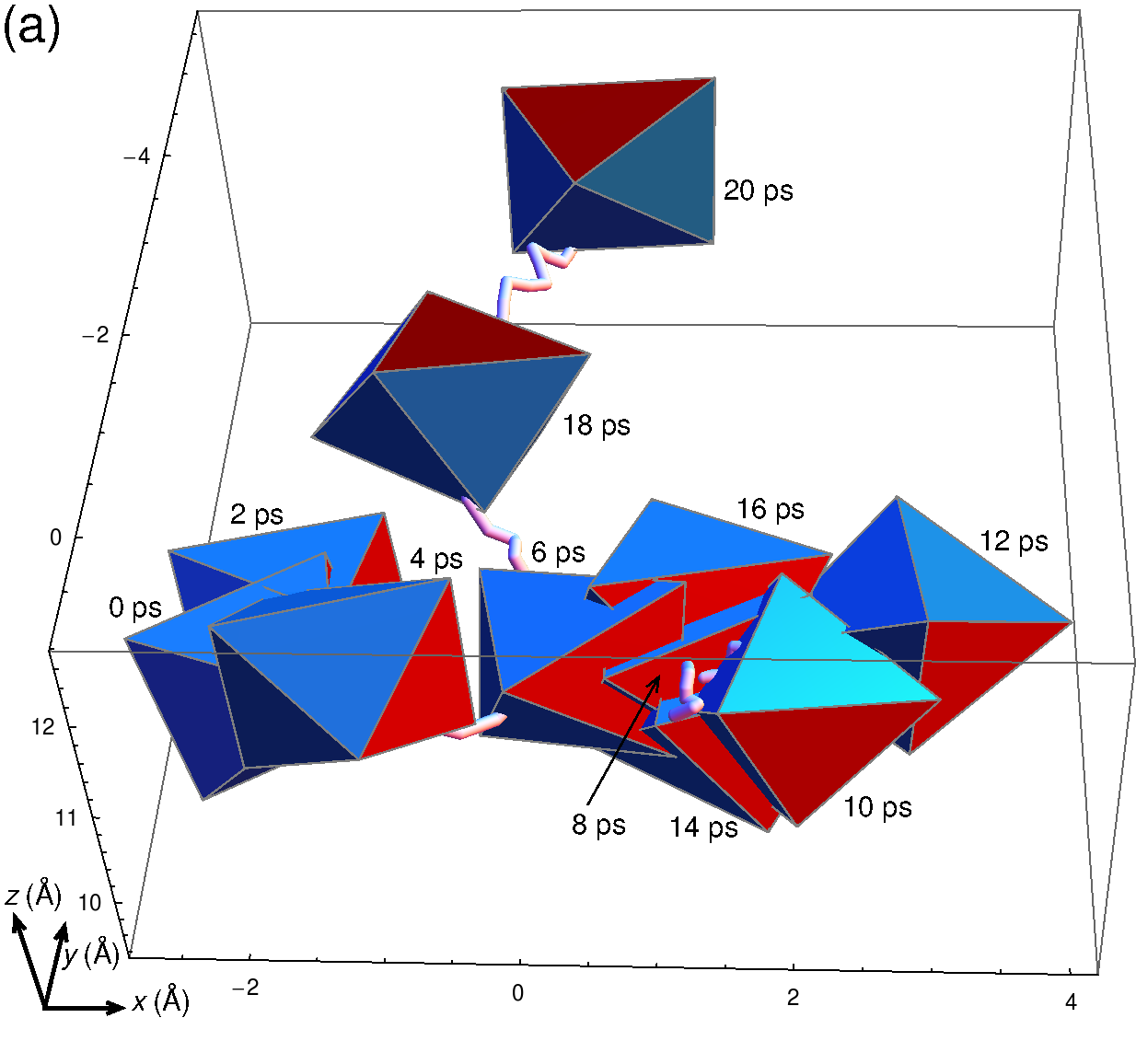}
\includegraphics[width=0.47\textwidth]{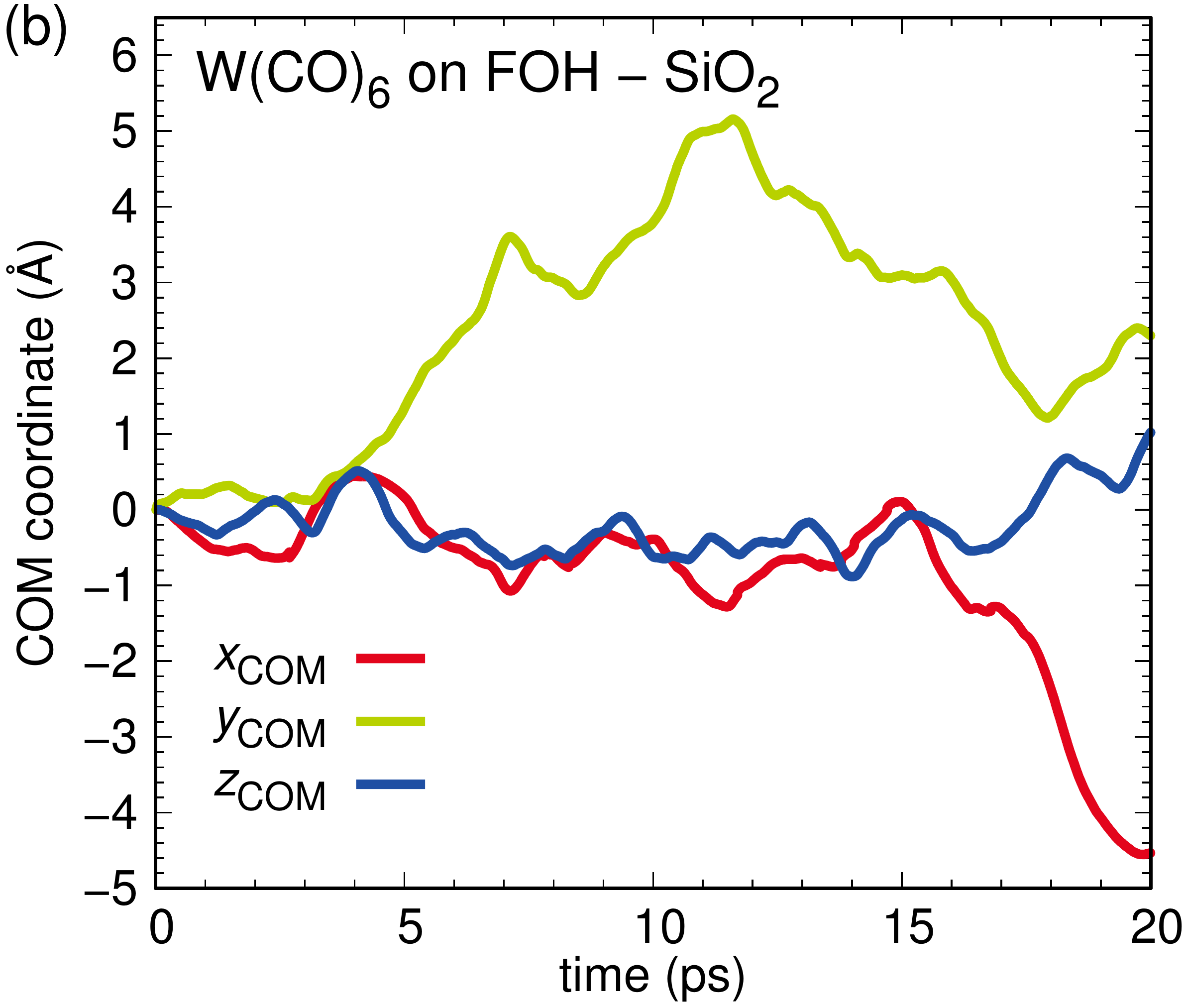}
\caption{(Color online) Schematic illustration of the movement of {\wc} on a {\FOH}
  surface. In the top panel, the polygon of the oxygen atoms of {\wc}
  is shown, with one facet always shown in red in order to visualize
  the rotations of the molecule. In the lower panel the time evolution
  of the three center-of-mass (COM) coordinates of {\wc} is shown. FOH in the figure
  corresponds to the fully hydroxylated surface}
\label{fig:wco6octahedra}
\end{figure}
%%%%%%%%%%%%%%%%%%%%%%%%%%%%%%%%%%%%%%%%%%%%%%%%%%%%%%%%%%%%%%%%%%%%%%%%%%%%%%%%%%%%%%%%%%%%%%%%%%%%%%%%%%%%%%%%%%%%

%%%%%%%%%%%%%%%%%%%%%%%%%%%%%%%%%%%%%%%%%%%%%%%%%%%%%%%%%%%%%%%%%%%%%%%%%%%%%%%%%%%%%%%%%%%%%%%%%%%%%%%%%%%%%%%%%%%%
\begin{figure}
\centering
\includegraphics[width=0.47\textwidth]{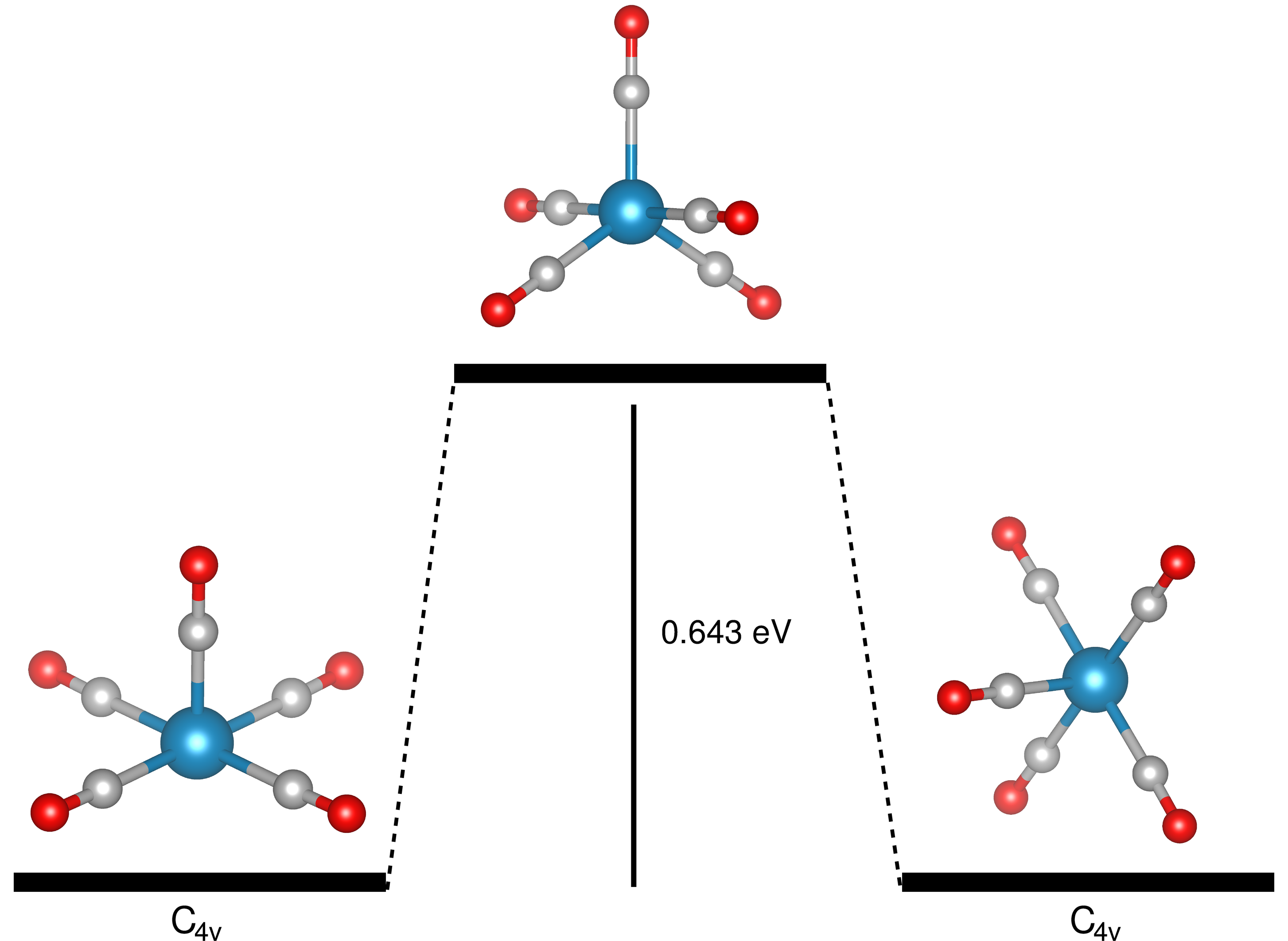}
\caption{(Color online) DFT optimized structure of \wcfive.  In the gas phase it
  possess a {\cfourv} symmetric structure.  The trigonal bipyramidal
  structure (middle panel) is 0.643~eV less stable than the square
  pyramidal structure (left and right panels)\cite{yishikowa1} but it
  is stabilized by surface-molecule interaction on \POH.  }
\label{fig:lessstable}
\end{figure}
%%%%%%%%%%%%%%%%%%%%%%%%%%%%%%%%%%%%%%%%%%%%%%%%%%%%%%%%%%%%%%%%%%%%%%%%%%%%%%%%%%%%%%%%%%%%%%%%%%%%%%%%%%%%%%%%%%%%

%%%%%%%%%%%%%%%%%%%%%%%%%%%%%%%%%%%%%%%%%%%%%%%%%%%%%%%%%%%%%%%%%%%%%%%%%%%%%%%%%%%%%%%%%%%%%%%%%%%%%%%%%%%%%%%%%%%%
\begin{figure}
\centering
\includegraphics[width=0.47\textwidth]{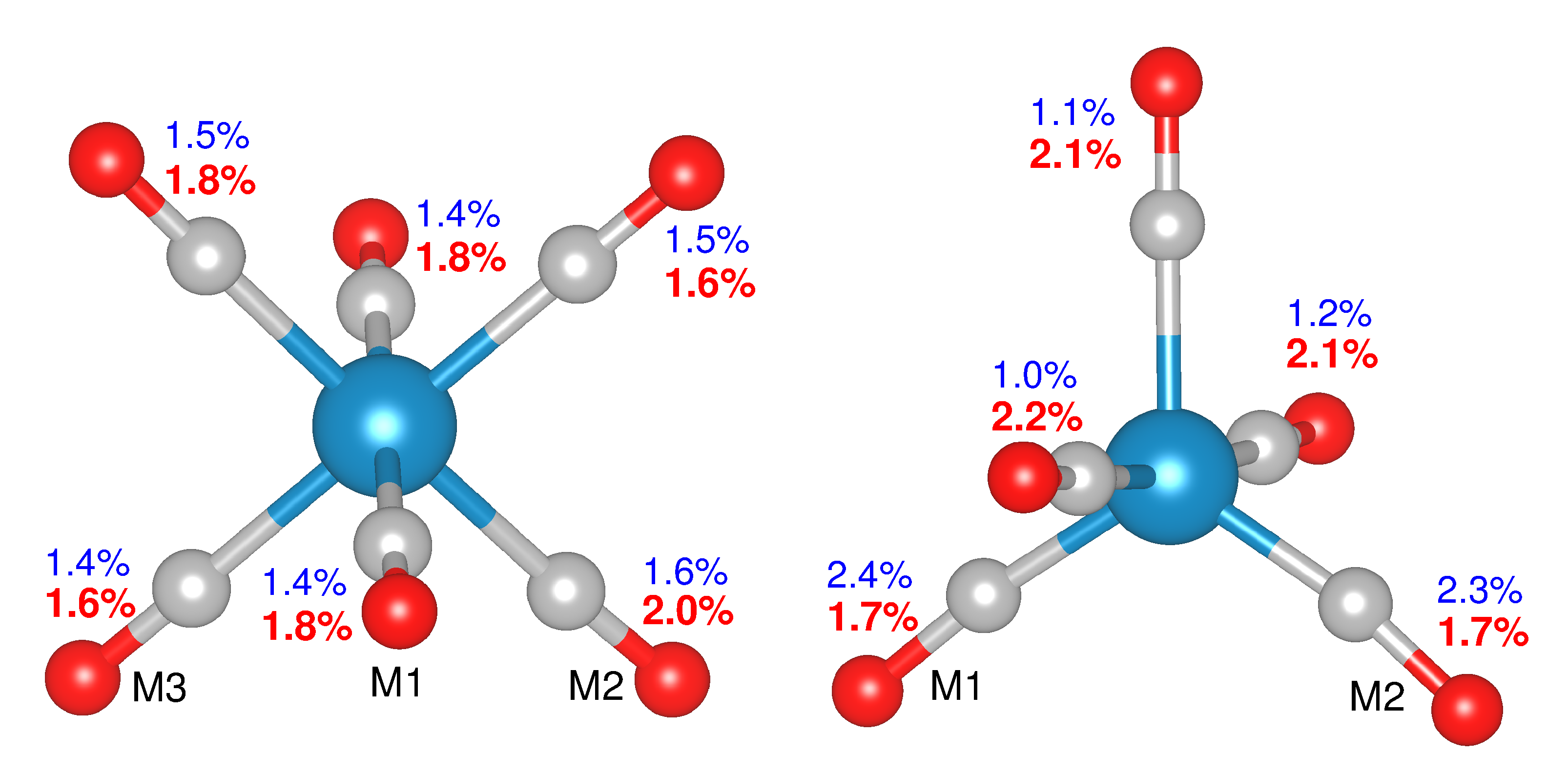}
\caption{(Color online) Variations in the structural features of {\wc} on {\sio}
  surfaces on {\FOH} (left panel) and {\POH} (right panel).  Bold
  (red) letters correspond to WC and normal letters (blue) for CO
  bond.  The values shown in this figure are the standard deviation of
  the respective values in the initial structure.  }
\label{fig:Standard_Deviation}
\end{figure}
%%%%%%%%%%%%%%%%%%%%%%%%%%%%%%%%%%%%%%%%%%%%%%%%%%%%%%%%%%%%%%%%%%%%%%%%%%%%%%%%%%%%%%%%%%%%%%%%%%%%%%%%%%%%%%%%%%%%

%%%%%%%%%%%%%%%%%%%%%%%%%%%%%%%%%%%%%%%%%%%%%%%%%%%%%%%%%%%%%%%%%%%%%%%%%%%%%%%%%%%%%%%%%%%%%%%%%%%%%%%%%%%%%%%%%%%%
\begin{figure*}
\includegraphics[width=1\textwidth]{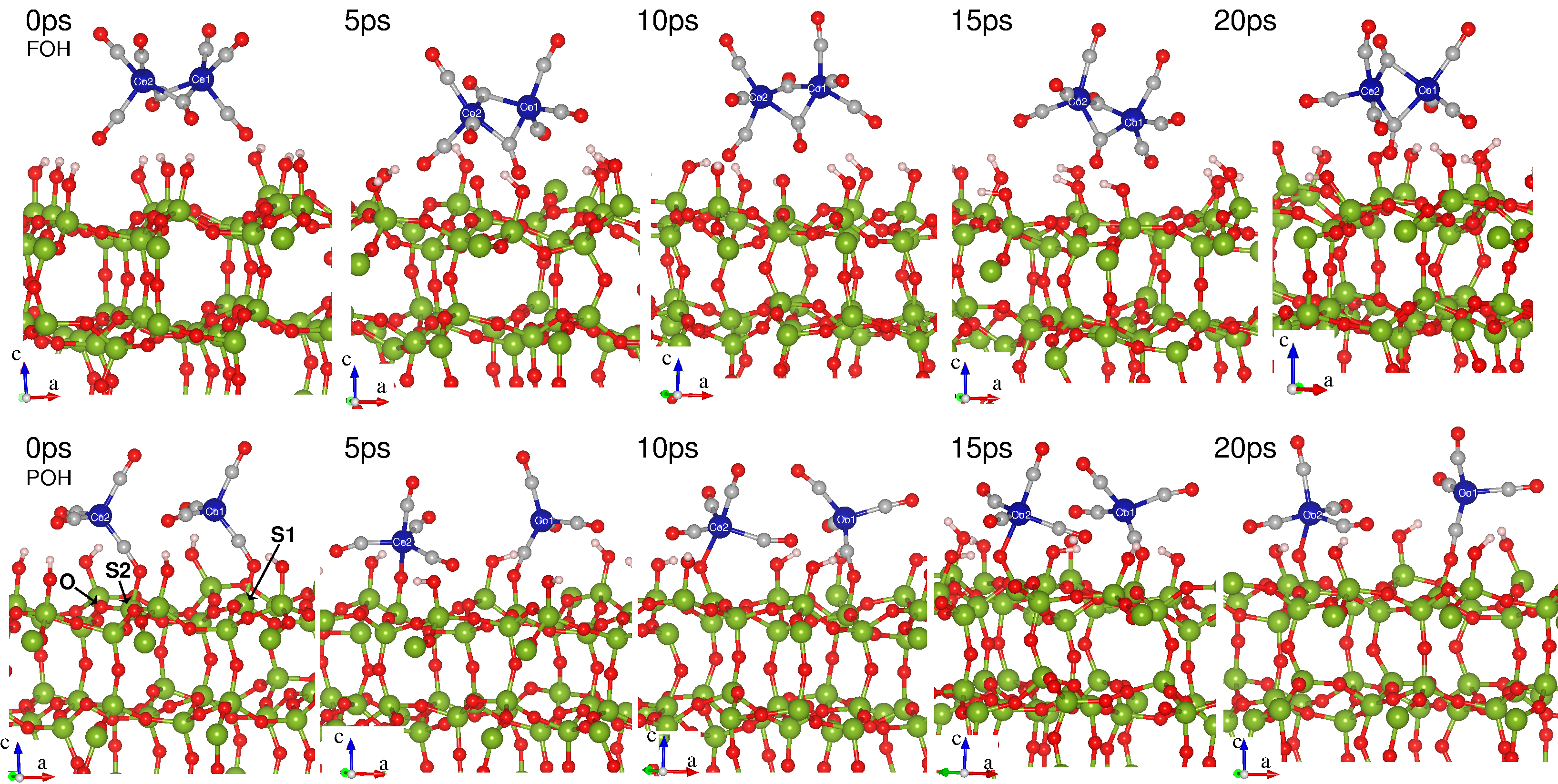}
\caption{(Color online) Adsorption of {\coco} on {\FOH} (top) and {\POH} (POH)
  substrates. The illustrated snapshots are the configurations taken
  at every 5ps interval.  S1 and S2 in {\POH} case denotes the two
  bonding sites on the surface where {\cofour} fragments
  are bonded.  Color code: Blue - Co, throughout this manuscript.
  The orientation of the configuration is denoted by the coordinate system on the bottom left hand side of each panel.}
\label{fig:cocosnapshot}
\end{figure*}
%%%%%%%%%%%%%%%%%%%%%%%%%%%%%%%%%%%%%%%%%%%%%%%%%%%%%%%%%%%%%%%%%%%%%%%%%%%%%%%%%%%%%%%%%%%%%%%%%%%%%%%%%%%%%%%%%%%%

%%%%%%%%%%%%%%%%%%%%%%%%%%%%%%%%%%%%%%%%%%%%%%%%%%%%%%%%%%%%%%%%%%%%%%%%%%%%%%%%%%%%%%%%%%%%%%%%%%%%%%%%%%%%%%%%%%%%
\begin{figure}
\centering
\includegraphics[width=0.47\textwidth]{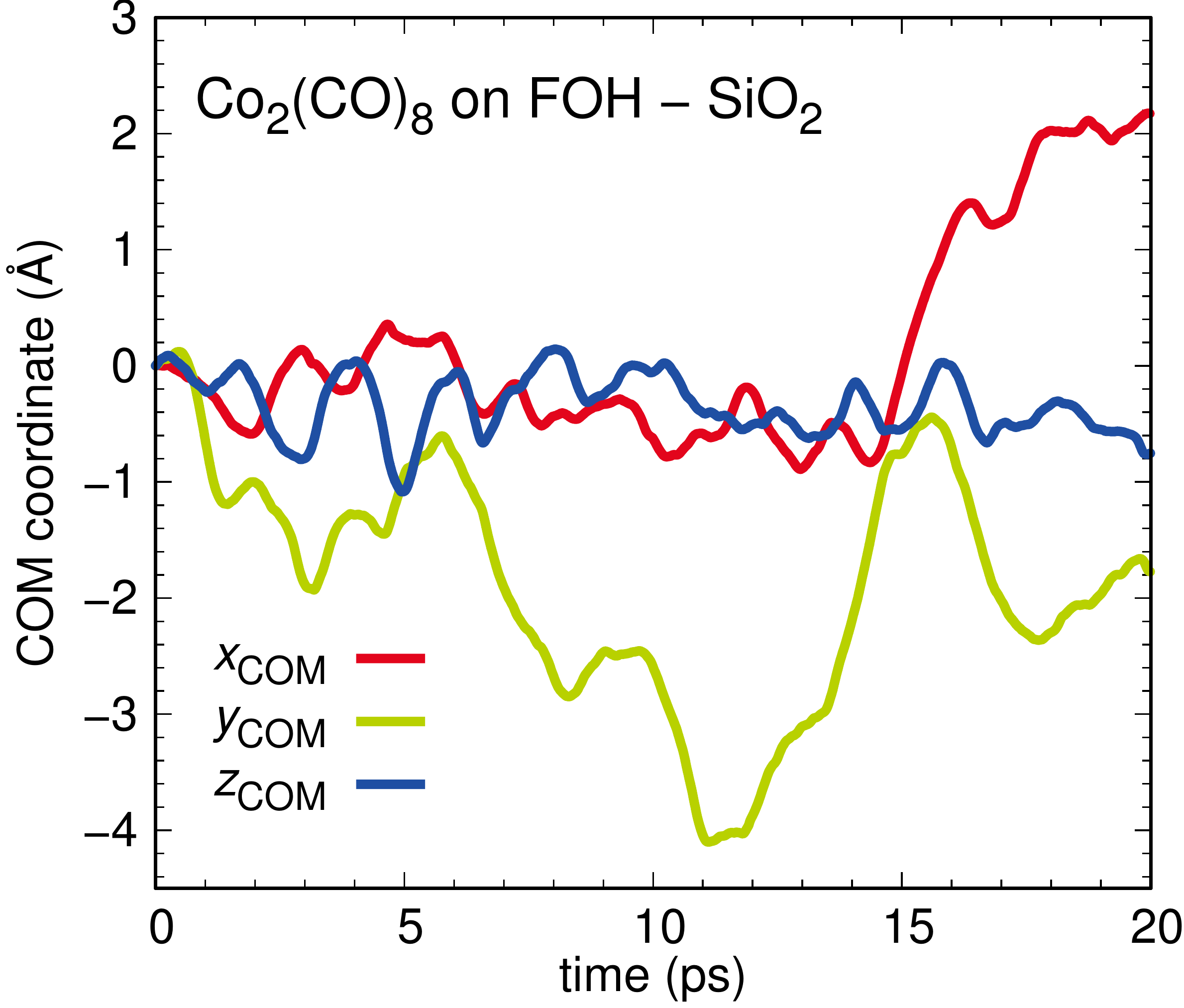}
\caption{(Color online) Evolution of the three center-of-mass coordinates of {\coco}
  adsorbed on a {\FOH} surface during the 20~ps trajectory.}
\label{cocoCM}
\end{figure}
%%%%%%%%%%%%%%%%%%%%%%%%%%%%%%%%%%%%%%%%%%%%%%%%%%%%%%%%%%%%%%%%%%%%%%%%%%%%%%%%%%%%%%%%%%%%%%%%%%%%%%%%%%%%%%%%%%%%

%%%%%%%%%%%%%%%%%%%%%%%%%%%%%%%%%%%%%%%%%%%%%%%%%%%%%%%%%%%%%%%%%%%%%%%%%%%%%%%%%%%%%%%%%%%%%%%%%%%%%%%%%%%%%%%%%%%%
\begin{figure}[htb]
\centering
\includegraphics[width=0.47\textwidth]{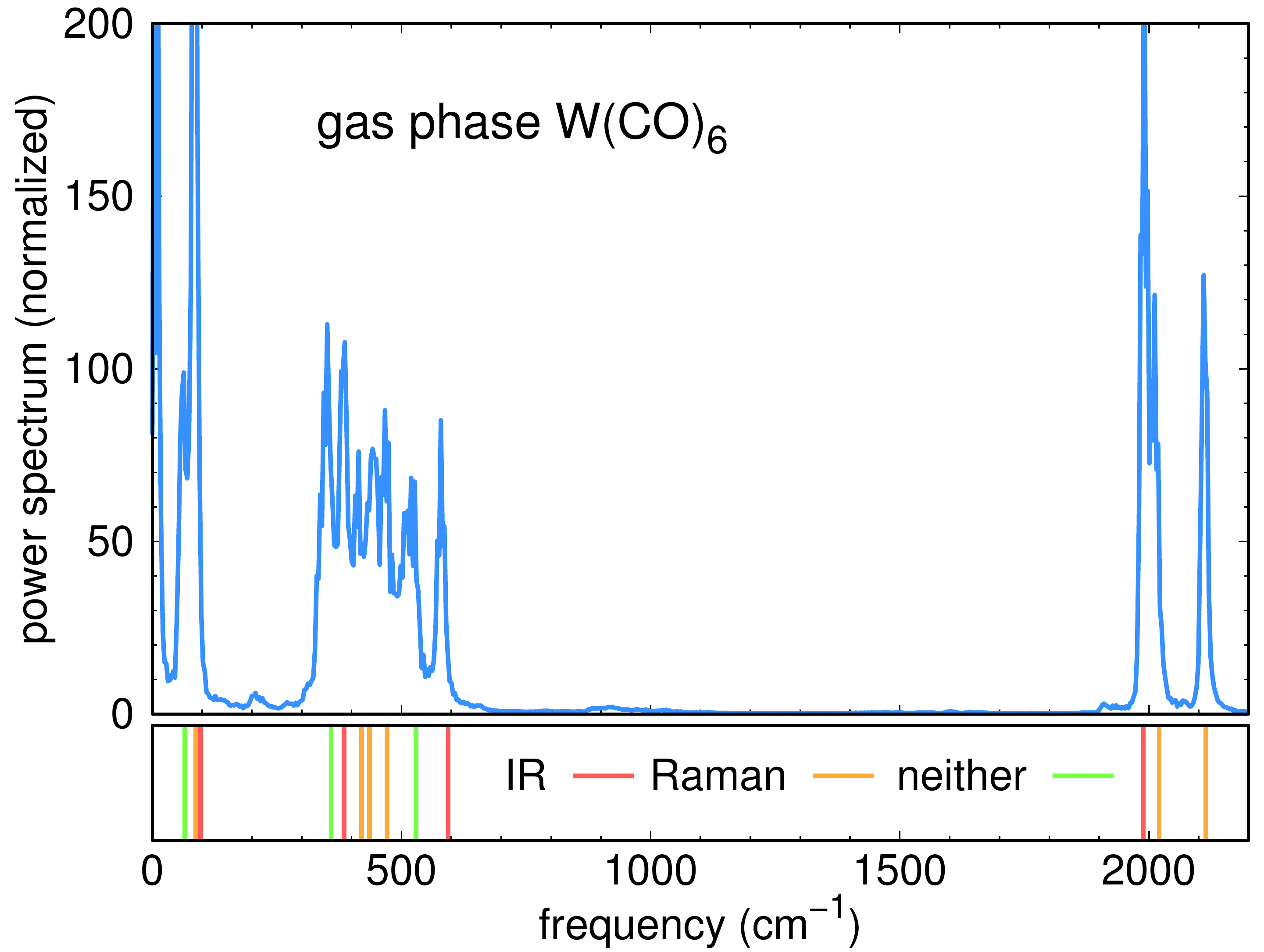}
\caption{(Color online) Calculated vibrational spectrum of the gas phase of {\wc}
  molecules by Fourier transforming the velocity autocorrelation
  function obtained from the MD trajectory (upper panel) and by the
  finite displacement method (lower panel).}
\label{fig:vib_gas}
\end{figure}
%%%%%%%%%%%%%%%%%%%%%%%%%%%%%%%%%%%%%%%%%%%%%%%%%%%%%%%%%%%%%%%%%%%%%%%%%%%%%%%%%%%%%%%%%%%%%%%%%%%%%%%%%%%%%%%%%%%%

%%%%%%%%%%%%%%%%%%%%%%%%%%%%%%%%%%%%%%%%%%%%%%%%%%%%%%%%%%%%%%%%%%%%%%%%%%%%%%%%%%%%%%%%%%%%%%%%%%%%%%%%%%%%%%%%%%%%
\begin{figure}[htb]
\centering
\includegraphics[width=0.47\textwidth]{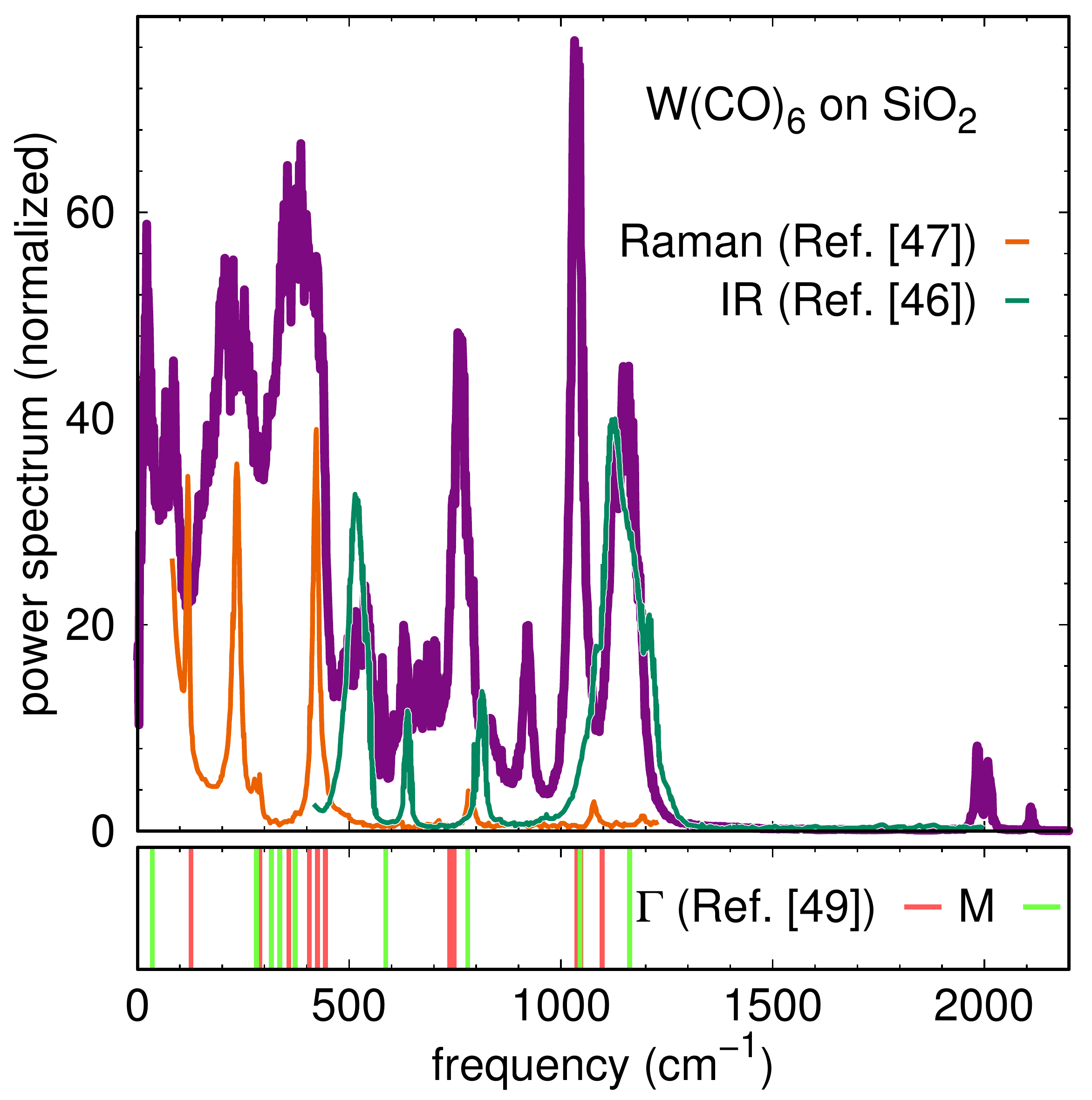}
\caption{(Color online) Comparison of the calculated power spectrum
  for {\wc} adsorbed on {\FOH} with the experimentally measured
  infrared and Raman spectrum~\cite{Finnie199423,bates:4042} for the
  {\sio} substrate. In the lower panel, the vibrational frequences of
  $\beta$ cristobalite at $\Gamma$ and M points in the Brillouin zone
  as calculated in Ref.~\cite{PhysRevB.78.054117} are shown.}
\label{fig:vib_exp}
\end{figure}
%%%%%%%%%%%%%%%%%%%%%%%%%%%%%%%%%%%%%%%%%%%%%%%%%%%%%%%%%%%%%%%%%%%%%%%%%%%%%%%%%%%%%%%%%%%%%%%%%%%%%%%%%%%%%%%%%%%%

%%%%%%%%%%%%%%%%%%%%%%%%%%%%%%%%%%%%%%%%%%%%%%%%%%%%%%%%%%%%%%%%%%%%%%%%%%%%%%%%%%%%%%%%%%%%%%%%%%%%%%%%%%%%%%%%%%%%
\begin{figure}[htb]
\centering
\includegraphics[width=0.47\textwidth]{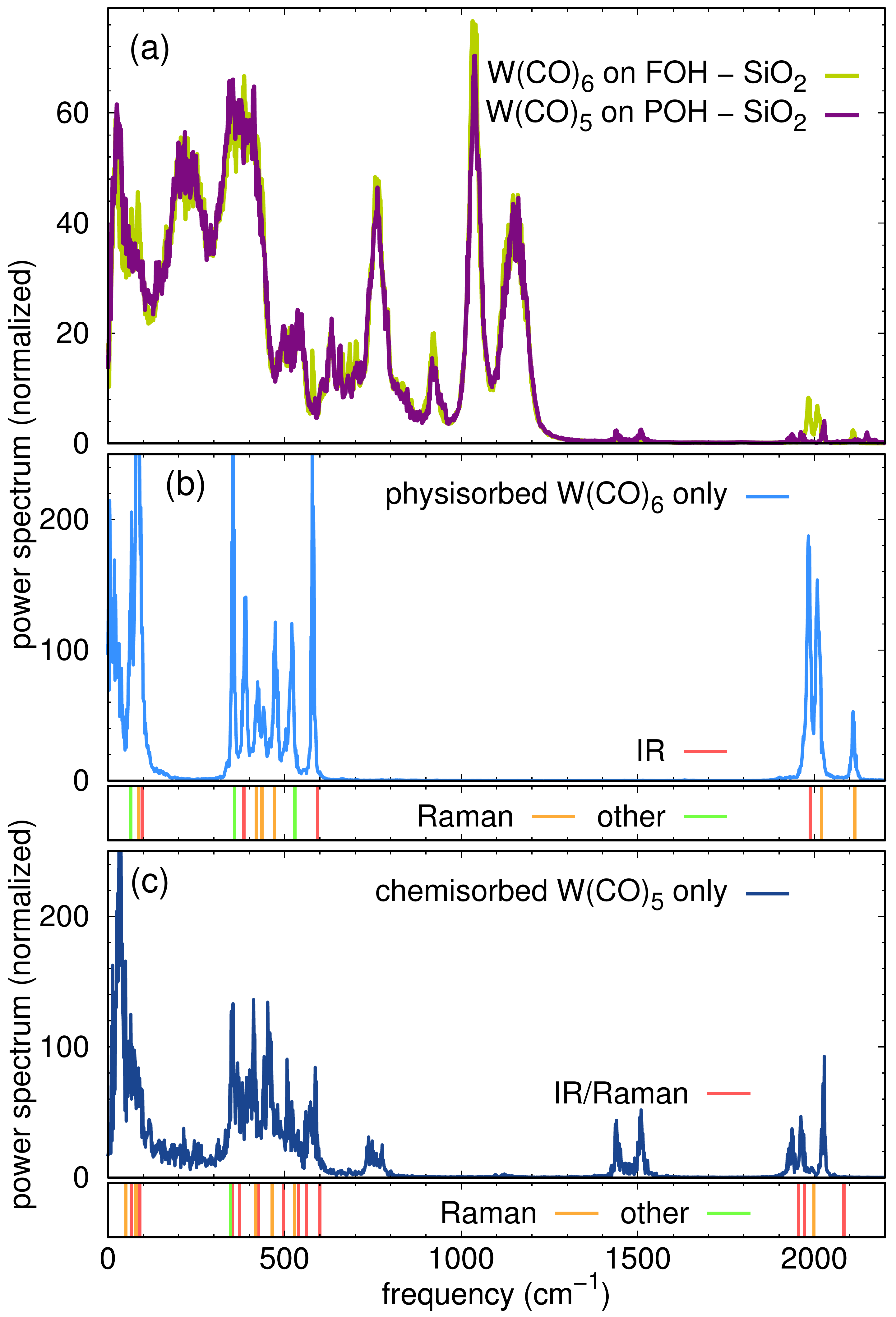}
\caption{(Color online) (a) Comparison of the total power spectrum for
  (i) {\wc} on a {\FOH} surface and (ii) {\wcfive} on a {\POH}
  surface. (b) Power spectrum of the physisorbed molecule {\wc} only
  in case (i) compared to the vibrational modes of {\wc} in the gas
  phase. The modes are classified in Raman, infrared and other. (c)
  Power spectrum of the chemisorbed molecule {\wcfive} only in case
  (ii) compared to the vibrational modes of {\wcfive} in the gas
  phase.}
\label{fig:vib_compare}
\end{figure}
%%%%%%%%%%%%%%%%%%%%%%%%%%%%%%%%%%%%%%%%%%%%%%%%%%%%%%%%%%%%%%%%%%%%%%%%%%%%%%%%%%%%%%%%%%%%%%%%%%%%%%%%%%%%%%%%%%%%

%%%%%%%%%%%%%%%%%%%%%%%%%%%%%%%%%%%%%%%%%%%%%%%%%%%%%%%%%%%%%%%%%%%%%%%%%%%%%%%%%%%%%%%%%%%%%%%%%%%%%%%%%%%%%%%%%%%%
\begin{figure}[htb]
\centering
\includegraphics[width=0.47\textwidth]{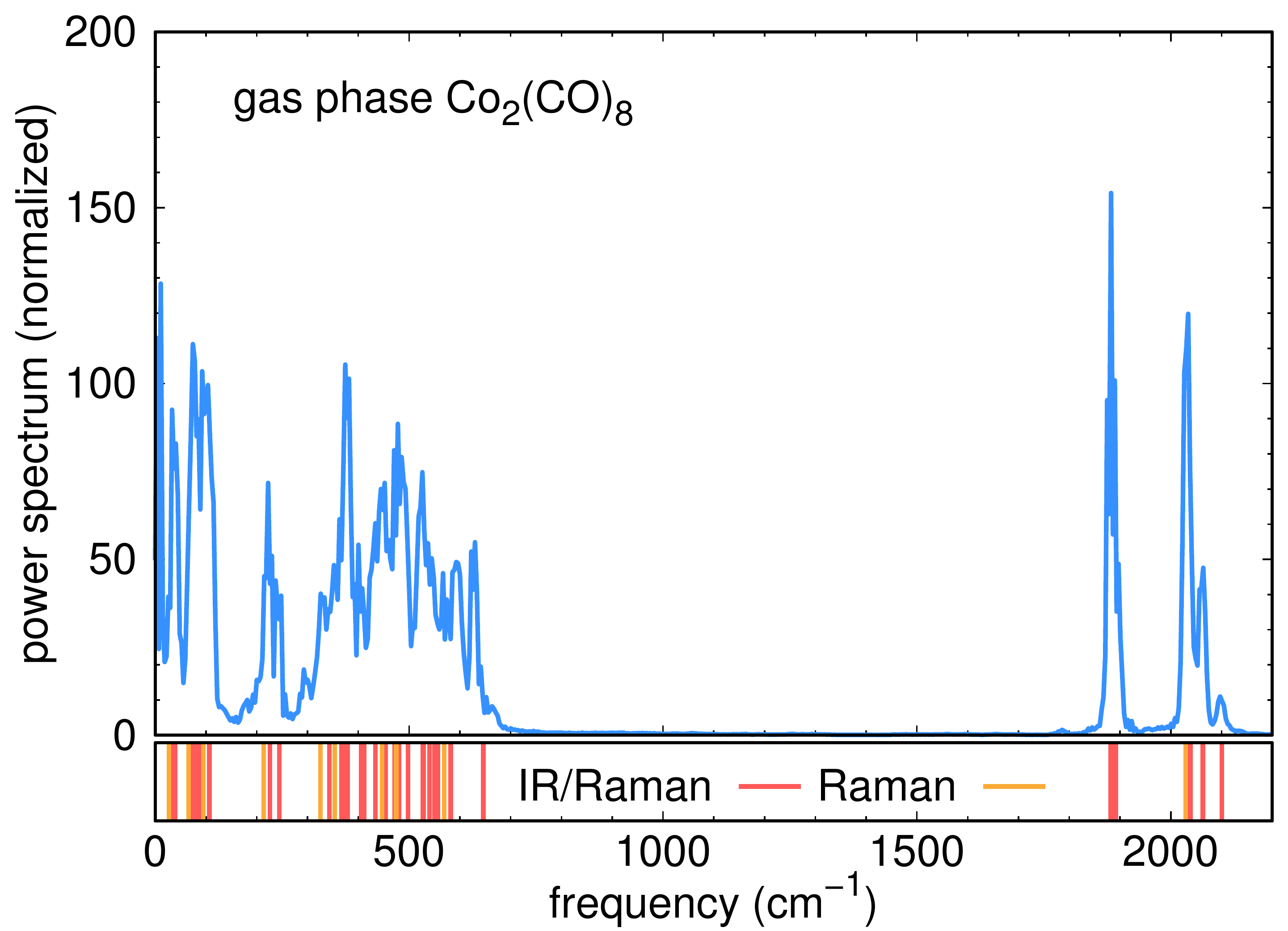}
\caption{(Color online) Calculated vibrational spectrum of the gas phase of {\coco}
  molecules by fourier transforming the velocities obtained from the
  MD trajectory (upper panel) and by the finite displacement method
  (lower panel). }
\label{fig:vib_gas_co}
\end{figure}
%%%%%%%%%%%%%%%%%%%%%%%%%%%%%%%%%%%%%%%%%%%%%%%%%%%%%%%%%%%%%%%%%%%%%%%%%%%%%%%%%%%%%%%%%%%%%%%%%%%%%%%%%%%%%%%%%%%%%

%%%%%%%%%%%%%%%%%%%%%%%%%%%%%%%%%%%%%%%%%%%%%%%%%%%%%%%%%%%%%%%%%%%%%%%%%%%%%%%%%%%%%%%%%%%%%%%%%%%%%%%%%%%%%%%%%%%%%
\begin{figure}
\centering
\includegraphics[width=0.47\textwidth]{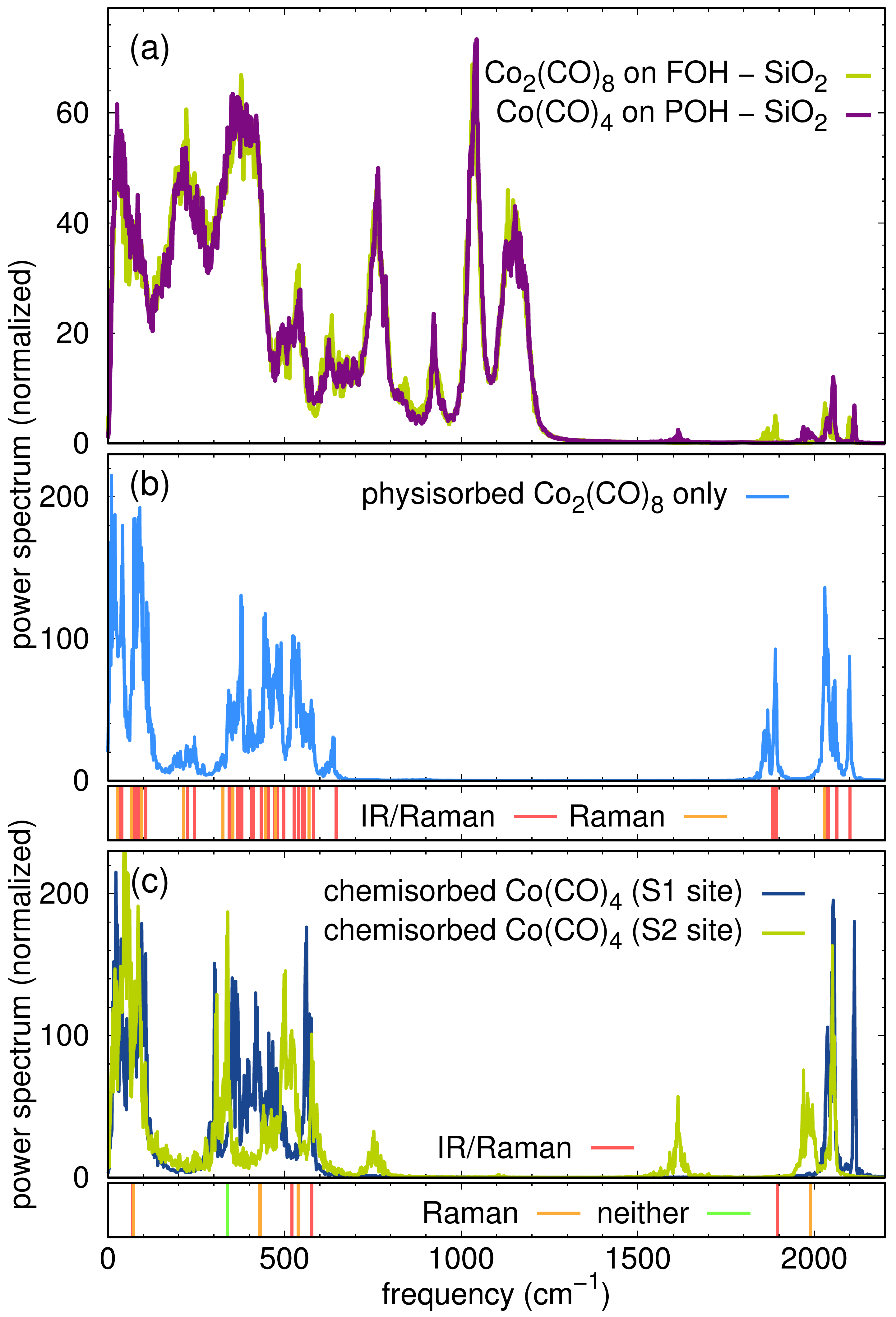}
\caption{(Color online) (a) Comparison of the total power spectrum for
  (i) {\coco} on a {\FOH} surface and (ii) the two {\cofour} fragments
  on a {\POH} surface. (b) Power spectrum of the physisorbed molecule
  {\coco} in case (i) compared to the vibrational modes of {\coco} in
  the gas phase.  (c) Power spectrum of the chemisorbed {\cofour}
  fragments in case (ii) compared to the vibrational modes of
  {\cofour} in the gas phase.}
\label{fig:vib_compare_Co}
\end{figure}
%%%%%%%%%%%%%%%%%%%%%%%%%%%%%%%%%%%%%%%%%%%%%%%%%%%%%%%%%%%%%%%%%%%%%%%%%%%%%%%%%%%%%%%%%%%%%%%%%%%%%%%%%%%%%%%%%%%%%%%

%%%%%%%%%%%%%%%%%%%%%%%%%%%%%%%%%%%%%%%%%%%%%%%%%%%%%%%%%%%%%%%%%%%%%%%%%%%%%%%%%%%%%%%%%%%%%%%%%%%%%%%%%%%%%%%%%%%%
\section{Computational details\label{Computationaldetails}}
%%%%%%%%%%%%%%%%%%%%%%%%%%%%%%%%%%%%%%%%%%%%%%%%%%%%%%%%%%%%%%%%%%%%%%%%%%%%%%%%%%%%%%%%%%%%%%%%%%%%%%%%%%%%%%%%%%%%
Ab initio molecular dynamics (MD) simulations were performed in the
framework of density functional theory (DFT).  We employed the
projector augmented wave (PAW) basis~\cite{Bloechl1994,Kresse1999}
within the generalized gradient approximation in the parametrization
of Perdew, Burke and Ernzerhof (PBE) as implemented in
VASP~\cite{Kresse1999,Kresse1993,Kresse1996a,Kresse1996b}.  Spin
polarized calculations including the corrections for long range van
der Waals interactions [37,38] were used in all calculations. All ions
were relaxed using the conjugate gradient scheme until the forces were
reduced by $\lesssim 0.01$~eV/{\AA} with a plane wave energy cut-off of
400~eV.  The Brillouin zone for the substrate-precursor complex was
sampled at the gamma point only.

The structures considered as initial configurations for the MD simulations are shown in Figs.~\ref{fig:snapshot} and \ref{fig:cocosnapshot} (see the $t=0$~ps snapshot).
MD simulations were performed for 20 ps on a
canonical ensemble at a finite temperature of $T=298$~K using the
Nose-Hoover thermostat.\cite{JCC:JCC21057} The temperature was chosen
in accordance with the reported experimental results where the largest
temperature rise during {\wc} deposition was $1\,^{\circ}{\rm C}$ when
a 1.42~nA electron beam was used for depositing tungsten
nanostructures.\cite{JJAP.46.6254,weber:461} The Verlet algorithm in
its velocity form with a time step of $\Delta t =1$~fs was used to
integrate the equations of motion. We also performed MD simulations for the
precursor molecules ({\wc} and {\coco}) in the gas phase by placing the molecules in a
cubic box of $a$ = $b$ = $c$ = 30~{\AA} using similar parameters used for the
substrate-precursor complex.
The length of the simulation for the molecule was limited to 10~ps.

The power spectrum I($\omega$) provides information about the
distribution of the vibrational energy of the system.
It can be computed by a Fourier transformation of the velocity autocorrelation function
obtained from the velocities ${\bf v}_i(t)$ of all atoms in the course
of the MD trajectory. The method we use is described in
the Appendix. Initially, we have compared the vibrations of the gas phase precursor molecules
obtained by the finite displacement method with the calculated power spectrum
(Vibrations through the finite displacement method were computed for the
precursor molecules {\wc} and {\coco} in the gas phase using Turbomole
6.0.\cite{treutler:346,Deglmann2004103,Eichkorn1995652} The geometries
were optimized using triple $\zeta$ basis sets for all elements and we
used an effective core potential (60 core electrons) for W in the case
of \wc.\cite{Eichkorn}). The vibrations computed by these two methods agree with each other and
are explained in Section \ref{vibrations}.  Therefore, vibrations for the \sio\ substrate and
the complex precursor molecule-substrate have been evaluated only from
the power spectrum so as to reduce the computational effort in
computing the Hessian matrix.

%%%%%%%%%%%%%%%%%%%%%%%%%%%%%%%%%%%%%%%%%%%%%%%%%%%%%%%%%%%%%%%%%%%%%%%%%%%%%%%%%%%%%%%%%%%%%%%%%%%%%%%%%%%%%%%%%%%%
\section{Results\label{Results and Discussion}}
%%%%%%%%%%%%%%%%%%%%%%%%%%%%%%%%%%%%%%%%%%%%%%%%%%%%%%%%%%%%%%%%%%%%%%%%%%%%%%%%%%%%%%%%%%%%%%%%%%%%%%%%%%%%%%%%%%%%
\subsection{Dynamics of {\wc} on {\sio} surfaces}
%%%%%%%%%%%%%%%%%%%%%%%%%%%%%%%%%%%%%%%%%%%%%%%%%%%%%%%%%%%%%%%%%%%%%%%%%%%%%%%%%%%%%%%%%%%%%%%%%%%%%%%%%%%%%%%%%%%%
Previous DFT calculations~\cite{PhysRevB.84.205442} indicate that
{\wc} interacts through weak physisorption with surface hydroxyls on
the {\FOH} surface and by strong chemisorption on the Si sites
available on {\POH} surfaces with substantial changes in the structure
and electronic properties.  The most stable configuration of {\wc} on
{\FOH} has an adsorption energy of -0.498~eV while the fragment
{\wcfive} together with a free CO ligand stabilize with an energy
-1.262~eV on a {\POH} surface.  In Fig.~\ref{fig:snapshot} we show
five snapshots in 5~ps intervals of the MD simulations of {\wc} on a
{\FOH} surface (upper panel) and of the fragment {\wcfive} together
with a free CO ligand on a {\POH} surface (lower panel).

Analysis of the trajectory in the {\FOH} case indicates that the {\wc}
molecule exhibits a considerable drift and moves around the initial
binding sites in a {\FOH} surface.  In order to visualize this
drifting, Fig.~\ref{fig:wco6octahedra} shows a schematic depiction of
the {\wc} displacement together with the center-of-mass (COM)
analysis.  The calculated drifting distance on the surface after 20~ps
is ca. 5~{\AA}.  These simulations illustrate that the undissociated
{\wc} molecule changes the orientation
considerably but does not desorb away from the {\FOH} surface.

However, a chemisorbed {\wcfive} molecule which is formed by the
release of a CO ligand from {\wc} (Fig.~\ref{fig:snapshot} lower
panel) remains localized on its binding site on a {\POH} surface.  The
dissociated CO ligand doesn't recombine with the parent moiety and the
vacant site on W remains empty and is not filled by surface hydroxyls
as has been suggested as a possibility for
{\mofive}.\cite{Reddyja00115a017} It should be noted that {\wcfive} in
the gas phase is stable on a square pyramidal structure and different
conformations are possible through the pathway shown in
Fig.~\ref{fig:lessstable} involving a trigonal bipyramidal transition
state.\cite{yishikowa2} Analysis of the adsorbed {\wcfive} structure (compare the transition
state on Fig.~\ref{fig:lessstable} with the configuration at 0~ps in
Fig.~\ref{fig:snapshot} lower panel) indicates that the {\wcfive}
molecule is stabilized in a trigonal bipyramidal structure on {\POH}
surfaces. The stabilization of such a transient intermediate has been
proposed for {\crfive} and the present work supports such a proposal
also for {\wcfive}. \cite{crcofive} This stabilization,
should have an impact over the kinetics of further release of intra
molecular CO ligands.

Evaluation of the changes in the WC and CO bond lengths of the
adsorbates {\wc} and {\wcfive} (see Fig.~\ref{fig:Standard_Deviation})
during the first 20~ps of the trajectory show deviations from the
initial configuration of the order of 1-2\% and might be due to
thermal fluctuations. We will analyze these deviations in more detail
in Section \ref{vibrations}.  The electronic structure of the final
configurations of the complex molecule-substrate on both fully and
{\POH} cases shows only minor variations with respect to the initial
configurations (results not shown).

%%%%%%%%%%%%%%%%%%%%%%%%%%%%%%%%%%%%%%%%%%%%%%%%%%%%%%%%%%%%%%%%%%%%%%%%%%%%%%%%%%%%%%%%%%%%%%%%%%%%%%%%%%%%%%%%%%%%
\subsection{Dynamics of {\coco} on {\sio} surfaces}
%%%%%%%%%%%%%%%%%%%%%%%%%%%%%%%%%%%%%%%%%%%%%%%%%%%%%%%%%%%%%%%%%%%%%%%%%%%%%%%%%%%%%%%%%%%%%%%%%%%%%%%%%%%%%%%%%%%%
We now proceed with the adsorption scenario of {\coco} interacting
with {\sio} surfaces. In Ref.~\cite{Muthukumar2012} a weak
bonding of {\coco} on {\FOH} surfaces was observed, similar to the
case of {\wc}, with an adsorption energy of -0.76~eV. In contrast,
{\coco} molecules on a {\POH} surface fragment into two {\cofour}
moieties rather than eliminating a CO ligand as in the case of
\wc.  The fragmented {\cofour} moieties exhibit a strong chemisorption
on {\POH} surfaces with an adsorption energy of -1.77~eV.\cite{Muthukumar2012}
Considering the above configurations as the initial setting for our MD simulations,
Fig.~\ref{fig:cocosnapshot} shows snapshots in 5~ps intervals of {\coco} adsorbed on a
{\FOH} surface (upper panel) and of {\cofour} fragments adsorbed on a {\POH}
surface (lower panel).

We observe a significant drift of {\coco} on the {\FOH} surface but as
this is a larger molecule than \wc, the displacement is less
pronounced than in \wc. This can be seen by comparing
Fig.~\ref{cocoCM}, where the COM movement of {\coco} on a
{\FOH} surface is depicted, with Fig.~\ref{fig:wco6octahedra}~(b)
which is analogous to {\wc}.  The calculated drifting radius of
{\coco} within 20~ps is about 4~{\AA}.

We now investigate the adsorption of the {\cofour} species on {\POH}
surfaces (Fig.~\ref{fig:cocosnapshot} lower panel).  We would like to
note that, unlike in the case of {\wcfive}, the {\cofour} species in
the gas phase possesses a tetrahedral structure which remains stable
with a slight distortion on the {\POH} surfaces (see $t=0$~ps in
Fig.~\ref{fig:cocosnapshot} lower panel). During the MD simulations,
the fragmented {\cofour} species are localized on the surface but we
observe severe changes in the orientation of CO ligands. In
particular, we find that the {\cofour} fragment bonded to the S2 site
on the {\POH} surface as shown in Fig.~\ref{fig:cocosnapshot} (lower
panel) shows bonding of the Co atom to surface oxygen as time evolves
(compare $t=0$~ps and $t=5$~ps snapshots).  A similar situation was
suggested by Rao {\it et al.}~\cite{Rao1988466} during the adsorption
of {\coco} on dehydroxylated MgO and on {\sio} surfaces.  However,
within our simulation window, we only observe this effect for the
{\cofour} fragment bonded to the S2 site but not for the fragment
bonded to the S1 site (see Fig.~\ref{fig:cocosnapshot} (lower panel))
of the partially hydroxylated {\sio} surfaces. Such a metal-substrate bond
might be due to activation of the substrate surface by
dehydroxylation.

%%%%%%%%%%%%%%%%%%%%%%%%%%%%%%%%%%%%%%%%%%%%%%%%%%%%%%%%%%%%%%%%%%%%%%%%%%%%%%%%%%%%%%%%%%%%%%%%%%%%%%%%%%%%%%%%%%%%
\subsection{Vibrations of {\wc} and {\coco} on a {\sio} surface \label{vibrations}}
%%%%%%%%%%%%%%%%%%%%%%%%%%%%%%%%%%%%%%%%%%%%%%%%%%%%%%%%%%%%%%%%%%%%%%%%%%%%%%%%%%%%%%%%%%%%%%%%%%%%%%%%%%%%%%%%%%%%
%%%%%%%%%%%%%%%%%%%%%%%%%%%%%%%%%%%%%%%%%%%%%%%%%%%%%%%%%%%%%%%%%%%%%%%%%%%%%%%%%%%%%%%%%%%%%%%%%%%%%%%%%%%%%%%%%%%%
\subsubsection{{\wc} adsorbed on {\sio} surface}
%%%%%%%%%%%%%%%%%%%%%%%%%%%%%%%%%%%%%%%%%%%%%%%%%%%%%%%%%%%%%%%%%%%%%%%%%%%%%%%%%%%%%%%%%%%%%%%%%%%%%%%%%%%%%%%%%%%%
We first start with the analysis of the vibrational spectrum of {\wc}
in the gas phase and the modes arising due to {\sio} surfaces.
We observe W$-$C stretching modes and (OC)$-$W$-$(CO) bending
modes  at 300-600~cm$^{-1}$, and C$-$O stretching modes at
1900-2150~cm$^{-1}$ for the gas phase {\wc}.
The bending vibrations that involve W$-$(CO) atoms are seen
at 60-100~cm$^{-1}$. All these normal modes are seen in both
upper (power spectrum) and lower panel (finite displacement method)
of Fig.~\ref{fig:vib_gas}. Also, in Fig.~\ref{fig:vib_exp}, we
compare the power spectrum calculated for {\wc} adsorbed on a {\FOH} surface with experimental infrared and
Raman spectra measured for {\sio} substrates so as to analyze the vibrations of {\sio}.~\cite{Finnie199423,bates:4042}
We observe a good overall agreement between the calculated
{\sio} modes (Infrared and Raman active peaks at 100-200, 400-450,
700-800 and 1000-1100~cm$^{-1}$) and the experimental
spectrum.\cite{liang:194524,PhysRevB.78.054117}
Apart from \sio\ modes, the additional modes in Fig.~\ref{fig:vib_exp}
are due to the surface adsorbed {\wc} molecule (compare with Fig.~\ref{fig:vib_gas}).
This analysis indicates that the adopted methodology
to calculate the power spectrum could be used to describe the vibrations of
{\wc} adsorbed on \sio\ surfaces.

In Fig.~\ref{fig:vib_compare} we present the total power spectrum
of {\wc} adsorbed on fully and {\POH} surfaces (Fig.~\ref{fig:vib_compare}(a))
and only the vibrations of physisorbed {\wc} and chemisorbed {\wcfive}
extracted from it which are compared with the vibrations (obtained by finite displacement method, shown as
vertical bars) of isolated gas phase structures (Fig.~\ref{fig:vib_compare}(b) and (c)).
From Fig.~\ref{fig:vib_compare} (b), we observe only minor changes in the vibrational modes
of {\wc}. In particular, the CO frequencies
observed between 1900 and 2110~cm$^{-1}$ (two of which are Raman active)
for the gas phase of {\wc} have also been observed
upon its bonding to {\FOH} surfaces without being perturbed
indicating that the molecule retains its gas-phase characteristics.
These features are indeed similar to {\wc} adsorbed on hydroxylated alumina ({\alumina})
surfaces where all the molecular vibrations are retained
\cite{doi:10.1021/ic50156a056,doi:10.1021/ic00274a022} indicating the
weak interaction between {\wc} and the substrate.

From the analysis of the chemisorbed {\wc} fragments on {\POH}
surfaces (see Fig.~\ref{fig:vib_compare}(c)) we find,
in addition to the shifts of several peaks, two new modes occurring at 700-800~cm$^{-1}$ and
1400-1600~cm$^{-1}$.
Peaks at 700-800~cm$^{-1}$ which correspond to the Raman vibrations of {\bc} \cite{liang:194524}, are
now seen, because of precursor bonding to the surface.
Also, the peaks at 1400-1600~cm$^{-1}$ appear as a result of oxygen (of CO)
interaction with substrate Si atoms.\cite{doi:10.1021/ic00274a022,Suvanto1999211}
Due to this interaction the bonding between tungsten and the carbonyl in {\wc}
gets affected leading to a considerable split of peaks in the
regions that correspond to the vibrations of C$-$O (1900-2100~cm$^{-1}$)
and W$-$C (300-600~cm$^{-1}$) bonds of {\wc} respectively.

%%%%%%%%%%%%%%%%%%%%%%%%%%%%%%%%%%%%%%%%%%%%%%%%%%%%%%%%%%%%%%%%%%%%%%%%%%%%%%%%%%%%%%%%%%%%%%%%%%%%%%%%%%%%%%%%%%%%
\subsubsection {{\coco} adsorbed on {\sio} surface}
%%%%%%%%%%%%%%%%%%%%%%%%%%%%%%%%%%%%%%%%%%%%%%%%%%%%%%%%%%%%%%%%%%%%%%%%%%%%%%%%%%%%%%%%%%%%%%%%%%%%%%%%%%%%%%%%%%%%
There have been some studies aimed at understanding the vibrations of {\coco}
adsorbed on {\sio}. However they are either limited to the carbonyl
region of the spectrum \cite{Rao1988466,10.1039/F19888402195} or they
observe immediate conversion of {\coco} to subcarbonyl species.\cite{doi:10.1021/ic00194a038,Suvanto:1999:0926-860X:25}
We first show in Fig.~\ref{fig:vib_gas_co} the analysis of the
vibrational spectrum of {\coco} in the gas phase
from power spectrum and the finite displacement method and both methods
agree with each other and the existing reports.\cite{doi:10.1021/ic001279n}
The two sets of vibrations in the region 1850-1900 and 2100-2000~cm$^{-1}$
are assigned to the stretching modes of the bridging and terminal CO
ligands and the Co-Co stretching mode in {\coco} is observed between
200-250~cm$^{-1}$.

We now investigate the vibrations of {\coco} adsorbed on {\sio} by
analyzing the power spectrum.
In Fig.~\ref{fig:vib_compare_Co} (a) we show the power spectrum of {\coco} on a {\FOH} surface
compared to the power spectrum of {\cofour} fragments on a {\POH} surface.
The extracted vibrations of physisorbed {\coco} and chemisorbed {\cofour}
from the total power spectrum are compared with the vibrations of
the isolated molecules and are shown in Fig.~\ref{fig:vib_compare_Co} (b) and (c) respectively.
The results illustrate that {\coco} on {\FOH} surface preserves
the gas-phase molecule characteristics (cf. Fig.~\ref{fig:vib_gas_co} and
Fig.~\ref{fig:vib_compare_Co} (b)).

In the case of {\cofour} on {\POH} surfaces, the overall agreement of the power
spectrum with that of the gas phase {\cofour} moiety is relatively
poor because the {\cofour} moieties exhibit strong structural
distortions upon bonding to the {\sio} substrate which is absent in
the gas phase computations. In the power spectrum, we find that the
modes in the region 220-250~cm$^{-1}$ (Co-Co stretching) and the mode at approx. 1850~cm $^{-1}$ (bridging carbonyls) disappear owing to the fact that the {\coco}
molecule fragments on {\POH} surfaces. Further, two new modes in the
600-800~cm$^{-1}$ and 1400-1600~cm$^{-1}$ regions appear in this case
(cf. Fig.~\ref{fig:vib_compare_Co} (a) and Fig.~\ref{fig:vib_compare_Co} (c)) illustrating the interaction of
surface Si atoms with the carbonyls through the oxygen atom of the
CO. However, this situation is only observed for one of the {\cofour}
moieties indicating that the fragmented species behave
differently. This is a consequence of the interaction of surface
oxygen atoms with the Co atom observed for the fragment bonded to the
S2 site (see Fig.~\ref{fig:cocosnapshot}).

%%%%%%%%%%%%%%%%%%%%%%%%%%%%%%%%%%%%%%%%%%%%%%%%%%%%%%%%%%%%%%%%%%%%%%%%%%%%%%%%%%%%%%%%%%%%%%%%%%%%%%%%%%%%%%%%%%%%
\subsection{Discussion}
%%%%%%%%%%%%%%%%%%%%%%%%%%%%%%%%%%%%%%%%%%%%%%%%%%%%%%%%%%%%%%%%%%%%%%%%%%%%%%%%%%%%%%%%%%%%%%%%%%%%%%%%%%%%%%%%%%%%
Investigation of surface adsorption and residence-time of the
precursors on the surface is necessary for improving the understanding
of deposition processes.  The present study elucidates the vibrational
footprints of precursors which interact through weak physisorption with
{\FOH} surfaces and by strong chemisorption with {\POH} surfaces.

The authors of Ref.~\cite{Domenichini200819} found that
physisorbed {\wc} molecules on {\tio} surfaces completely desorb when
the system is cooled down to room temperature.  Our study finds no
desorption of {\wc} molecules on {\FOH} surfaces in the considered
length of simulation.  Also, {\coco} molecules on {\FOH} surfaces were
reported to fragment spontaneously \cite{doi:10.1021/ic00194a038,doi:10.1080/01614949308014607}
but, in our calculations on {\FOH} surfaces we didn't observe any tendency
to fragmentation or an indication of chemisorption. Other effects
(probably extrinsic) may be responsible for the experimental
observations.

The bond variations of {\wc} and {\coco} adsorbed on {\FOH} surfaces
indicate uniform fluctuations (W$-$C and C$-$O bonds) on either
side (i.e, oriented towards the surface and the vacuum) of the molecule.
On the contrary, the relative bond values of {\wcfive} on
{\POH} surfaces show non-uniform variations indicating that certain
bonds (i.e., W$-$C bonds oriented towards the vacuum and the CO bonds
towards the substrate) experience larger changes than others.  Thus,
for favorable conditions, the bond between C and O for the CO bonded
to the surface might cleave leaving the surface Si atoms terminated
with oxygens. This situation is not observed in the present MD
simulations but the higher ratio of carbon contamination with respect
to oxygen (before atmospheric air exposure) in the EBID-obtained
samples might provide an evidence for such a
fact.\cite{1367-2630-11-3-033032,Botman20081139,li:023130,barry:3165,0957-4484-20-19-195301}
Similarly, we found in the case of {\coco} on {\POH} surfaces, that
surface oxygen atoms are involved in bonding to the dissociated
surface species. This suggests that the removal of oxygen components
from the deposits might not be an easy task to achieve.  Presence of
oxygen contamination is also expected to occur as a result of exposure
of EBID deposits to the air.  Thus the composition of the EBID
deposits are determined to a large extent by the number of available
active Si sites (alternatively the degree of hydroxylation), the
frequency of these precursor molecules approaching such a site and the
exposure time to the environments. Furthermore, the surface defects
observed in {\sio} during the interaction of {\cofour} with {\POH}
surfaces might also act as an active site for activating the
approaching precursor molecule that could account for the facile
dissociation of {\coco} on {\sio}.\cite{Muthukumar2012,co2frag}

On {\POH} surfaces our results for both {\wc} and {\coco} show that
the fragmented species remain localized, thus blocking active sites on
the surface.  Therefore, further deposition should occur on the
deposited layers and this may be likely the reason for the increase in
height of the deposits as the irradiation time increases during EBID
experiments.\cite{doi:10.1021/nn203134a}

%%%%%%%%%%%%%%%%%%%%%%%%%%%%%%%%%%%%%%%%%%%%%%%%%%%%%%%%%%%%%%%%%%%%%%%%%%%%%%%%%%%%%%%%%%%%%%%%%%%%%%%%%%%%%%%%%%%%
\section{Conclusions\label{Conclusion}}
%%%%%%%%%%%%%%%%%%%%%%%%%%%%%%%%%%%%%%%%%%%%%%%%%%%%%%%%%%%%%%%%%%%%%%%%%%%%%%%%%%%%%%%%%%%%%%%%%%%%%%%%%%%%%%%%%%%%
The purpose of this work was to model by means of {\it ab initio}
molecular dynamics simulations the dynamics of two precursor molecules
adsorbed on fully and partially hydroxylated {\sio} surfaces in order to achieve a better
understanding of the microscopics of electron-beam induced deposition
of nanostructures.  Our results reveal that {\wc} and {\coco}
molecules preserve their gas-phase bonding characteristics on {\FOH}
surfaces.  Apart from a considerable drift, only minor variations in
the structure and vibrations is observed.  Therefore spontaneous
dissociation of these precursor molecules will not be possible on
{\FOH} surfaces, unless some surface active sites are created by external
forces.

For the case of {\wc} and {\coco} on {\POH} surfaces, the fragmented
species retain the chemisorbed character on the surface and we do not
observe any reformation of the parent precursor moiety, but instead a
slight tendency towards fragmentation.  We also observe a smaller
weakening of the surface-oriented CO bonds compared to vacuum-oriented CO
bonds in {\wc} and {\coco} on {\POH} surfaces.  Therefore, conditions
that favor the formation of active sites (in this case surface Si
atoms) are needed in order to have high efficiency in fragmentation
and improve the metal content of the deposit.

The calculated vibrational spectra of these carbonyl molecule/ \sio\ substrate
systems show clear fingerprints to be detected experimentally. We propose therefore
the consideration of such simulations as a route to experimentally distinguish
the form in which precursors cover a substrate.

Finally,  while the present simulations provide insights on the surface-precursor interaction, the
investigation of other processes like surface-electron,
precursor-electron and  deposit-molecule interaction remains a
challenge for future work.

\section{Acknowledgments}

We would like to thank the Beilstein-Institut, Frankfurt/Main,
Germany, within the research collaboration NanoBiC for financial
support. The generous allotment of computer time by CSC-Frankfurt and
LOEWE-CSC is also gratefully acknowledged.
%Recently experimental and phenomenological
%studies focussing on these topics /have been reported \cite{RCM:RCM6324}
%and to study all these interactions through first principles calculations
%remains a challenge for future work.

\begin{appendix}
\section{Estimation of the power spectrum}\label{sec:power}
We use the power spectrum for analysing the vibrational
characteristics of precursor molecules in the gas phase and for the
entire system of precursors chemi- or physisorbed on a silica
substrate~\cite{diss}. The power spectrum is defined as~\cite{noid:404,chang:7354}
\begin{equation}
I(\omega) = \frac{1}{2 \pi}\int_{-\infty}^{\infty}C(\tau) e^{-i \omega \tau} d \tau\,,
\end{equation}
where $C(\tau)$ is the velocity autocorrelation function
\begin{equation}\label{eq:veloac}
C(\tau) = \langle \mathbf{v}(0)\mathbf{v}(\tau) \rangle = \lim_{T\to \infty} \left[
\frac{1}{T\,N}\int_0^T \sum_{l=1}^N \mathbf{v}^l(t)\mathbf{v}^l(t+\tau) dt
\right]\,.
\end{equation}
$\mathbf{v}^l(t)$ represents the velocity of atom $l$ at time $t$ for
all $N$ atoms of the system.  As the calculation of the time averages
$\langle \dots\rangle$ of Eq.~(\ref{eq:veloac}) is quite inefficient
computationally, we use the Wiener-Khinchin theorem which guarantees
that the power spectrum can also be calculated by individually
Fourier-transforming the velocities $\mathbf{v}^l(t)$ and summing the
squares of the result:
\begin{equation}\label{eq:wiener}
I(\omega) = \frac{1}{N} \sum_{l=1}^N \left| \left[\frac{1}{2 \pi}\int_{-\infty}^{\infty}\mathbf{v}^l(t)
e^{-i \omega t} d t \right] \right|^2\,.
\end{equation}
Finite trajectories,
calculated with a finite time step $\Delta t$, only yield estimates to the power
spectrum. Instead of the Fourier integral in
Eq.~(\ref{eq:wiener}), discrete Fourier sums have to be calculated
according to
\begin{equation}\label{eq:fourier}
V^{l \mu}_k = \sum_{j=0}^{J-1} v_j^{l \mu} e^{\frac{2 \pi i j k}{J}}\,,
\end{equation}
where the $v_j^{l \mu}$ stand for the $\mu$ component ($\mu \in \{x,y,z\}$) of
the velocity of atom $l$ at time step $j$, and $J$ is the number of
time steps of the trajectory. The so-called periodogram estimate for
the power spectrum is then defined for $J/2+1$
frequencies~\cite{Press}:
\begin{equation}\begin{split}\label{eq:periodogram}
P^{l \mu}(0) = P^{l \mu}(f_0) =& \frac{1}{J^2}|V^{l \mu} _0|^2  \,,\\
P^{l \mu}(f_k)  =& \frac{1}{J^2}\Bigl[|V^{l \mu} _k|^2+|V^{l \mu} _{J-k}|^2\Bigr] \,,\\
P^{l \mu}(f_c) = P^{l \mu}(f_{J/2}) =& \frac{1}{J^2}|V^{l \mu} _{J/2}|^2 \,.
\end{split}\end{equation}
The highest frequency $f_c = \frac{1}{2\Delta t}$ is called Nyquist
frequency and is determined by the time step $\Delta t$ of the MD
calculation. Thus, the finite time estimate of the power spectrum can
finally be written as
\begin{equation}\label{eq:wiener1}
I^J(\omega) = \frac{1}{N} \sum_{l=1}^N \sum_{\mu=1}^3  P^{l \mu}(\omega)\,.
\end{equation}
The function $I^J(\omega)$ will approach the true power spectrum
$I(\omega)$ of the system in the limit $J \to \infty$, {\it i. e.} in
the limit of infinitely long trajectories. The power spectrum
estimates now provide us with information about the distribution of
the vibrational energy of a molecule or a solid over the
frequencies. Power spectra at different temperatures may be used to
investigate the differences in the population of vibrational
modes. The integral over the power spectrum corresponds to the kinetic
energy of the system.
\end{appendix}

%\section{References}

\end{document}